\documentclass[aps,pra,preprintnumbers]{revtex4}
\usepackage{graphicx}
\usepackage{dcolumn}
\usepackage{bm}
\usepackage{hyperref}
\usepackage[latin1]{inputenc}
\usepackage{pstricks}           

\newcommand{\rs}{{\bf r}}

\newcommand{\vs}{{\bf v}}

\newcommand{\ec}{{\bf E}}

\newcommand{\eps}{{\bf \varepsilon}}

\begin{document}

\title{Field emission from metal surfaces in the Thomas-Fermi-von-Weizs\"{a}cker model}

\author{Boyan Obreshkov$^{1,2,3}$}

\affiliation{$^1$ Universit\'{e} de Toulouse, UPS, Laboratoire
Collisions Agr\'{e}gats R\'{e}activit\'{e}, IRSAMC, Toulouse,
France}

\affiliation{$^2$ CNRS, UMR 5589, F-31062, Toulouse, France}

\affiliation{$^3$Institute for Nuclear Research and Nuclear
Energy, Bulgarian Academy of Sciences, Tsarigradsko chausse\'{e}
72, Sofia 1784, Bulgaria}

\author{Bruno Lepetit$^{1,2}$}
\affiliation{$^1$ Universit\'{e} de Toulouse, UPS, Laboratoire
Collisions Agr\'{e}gats R\'{e}activit\'{e}, IRSAMC, Toulouse,
France}

\affiliation{$^2$ CNRS, UMR 5589, F-31062, Toulouse, France}

\date{\today}

\begin{abstract}

We evaluate the electron emission current density from jellium
metallic surfaces in the Thomas-Fermi-von-Weizs\"{a}cker
approximation. We implement the weighted density approximation
(WDA) for description of the exchange and correlation energy of
interacting electrons. We find electron-emission exhibits
crossover from quantum to classical over-barrier escape of
electrons from the surface barrier. Well below a surface-specific
threshold field strength $E_d$, electron tunneling is mainly
affected by the change of the metal workfunction, and is less
sensitive to the detailed shape of the surface barrier. In
particular, since the position of the image charge plane for
electrons leaving the surface does not precisely coincide with the
centroid of the induced screening charge at the surface in
response to the applied electrostatic field, we find an effective
increase in the metal workfunction by $\Delta W \sim 0.01$ eV,
which decreases the dark current.

\end{abstract}

\maketitle


\section{Introduction}

Metallic surfaces subject to strong electric field emit electrons
into vacuum due tunneling transitions \cite{FN_orig}.  Employing
the ballistic motion of emitted electrons in vacuum, field
emission has found important practical applications in development
of vacuum microelectronic devices \cite{fursey}, at the same time
electron emission is an unwanted phenomenon in high-voltage vacuum
insulation where it is considered as precursor to breakdown
\cite{Latham,hviv}. Besides its technological importance and
experimental interest, there remain relatively few calculations of
the electron emission current using first-principles methods for
description of the field-dsitroted surface electronic structure
\cite{Gohda,Liebsch,steps}. For the field emission from high work
function surfaces Ref.\cite{Liebsch,steps}, the calculated
electron current density $J$ complies qualitatively the
Fowler-Nordheim (FN) model \cite{FN_orig}, which obeys a simple
dependence upon the applied electric field $E$, i.e.
\begin{equation}
J_{{\rm FN}}(E)= \sqrt{\frac{\epsilon_F}{W}}
\frac{1}{W+\epsilon_F} E^2 \exp\left[-\frac{4}{3} \left(\frac{2
m}{\hbar^2} \right)^{1/2} \frac{W^{3/2}}{e E}\right], \label{FN}
\end{equation}
where $W$ is the substrate workfunction, $\epsilon_F$ designates
the position of the Fermi level relative to the lower edge of the
valence band, $h=2 \pi \hbar$ is the Planck constant and $-e$ and
$m$ are the charge and mass of an electron, respectively.

Common to such model calculations is the use of an effective
one-electron potential energy appropriate for vacuum tunneling,
which is obtained in the framework of density functional theory
(DFT) using the local density approximation (LDA) for the exchange
and correlation energy. A well known drawback of LDA is that it
fails to describe classical long-range image charge interaction
between electron outside the surface and the polarization charge
density that builds up at the surface in response to the Coulomb
field of the escaping electron. Image charge interactions are also
known to affect measured conductance of tunneling barriers arising
in gallium-arsenide structures \cite{expt1_image} and play an
important role in the correct interpretation of STM measurements
\cite{binnig_image}.

Nordheim \cite{N28} extended the FN model by including the Coulomb
interaction energy $U_{{\rm im}}(z)=-e^2/16 \pi \eps_0 z$ of an
electron at a distance $z$ from the surface with its induced
self-image charge, where $\eps_0$ is the vacuum permittivity
constant. This was discussed further in \cite{image1} and
elaborated by Murphy and Good \cite{FN_image}, the result for the
current density including the image potential is
\begin{equation}
J_{{\rm MG}}(E)= \frac{-e}{8 \pi h W t^2(y)} (eE)^2 \exp
\left[-\frac{4}{3} \left( \frac{2m}{\hbar^2}
\right)^{1/2}\frac{W^{3/2}}{e E} v(y) \right], \label{cfe}
\end{equation}
where $v(y)$ at $t(y)$ are slowly varying functions of the
Nordheim parameter $y=(e^3 E/4 \pi \eps_0)^{1/2}/W$, and are
related to certain linear combination of elliptic functions. Thus
classical image charge effects take the form of multiplicative
correction to the Fowler-Nordheim exponent. Normally it is assumed
that Eq.\ref{cfe} remains valid for more realistic
electron-surface potentials, because the main effect of reduction
of the surface barrier occurs at distances where the classical
image potential provides reasonable approximation. Thus
Eq.(\ref{cfe}) is used in the interpretation and analysis of
experimental electron-emission data in terms of Fowler-Nordheim
plot of $\ln(J/E^2)$ as a function of the inverse field strength
$1/E$. Field enhancement factor is extracted from the slope and an
effective emitting area from the intercept of the FN plot. These
parameters depend on the surface roughness, morphology and other
specific geometrical features of the emitting surface. To
investigate influence of the surface barrier shape, more refined
analysis of the FN plot has been developed \cite{Forbes}, where
slope and intercept correction factors are introduced in order to
obtain more reliable values of the emission area and the field
enhancement factor.

However Eq.(\ref{cfe}) does not describe correctly the
electron-surface interaction at small distances $z$, where the
classical image potential energy $U_{{\rm im}}(z)$ diverges.
Within DFT \cite {lk1} for a semi-infinite jellium model
describing metal surface, the one-electron potential energy
saturates to the local density exchange and correlation potential
energy \cite{lk2} close to the surface. Since electron tunneling
is sensitive to the change of this potential energy, the
motivation for the present work is development of more elaborate
model for the surface barrier for electrons, which allows to
quantify exchange and correlation effects in field emission from
simple metal surfaces. The problem is simplified when the time
spent by electrons under the barrier is long compared to the time
required for the surface screening charges to build up
\cite{time,BP}, such that dynamical screening effects may be
neglected and the use quasistatic approximation for image charge
interactions can be justified.

This paper is organized as follows. In Sec. II we give details on
our theoretical model for field emission from arbitrarily
structured metallic surfaces in the framework of density
functional theory (DFT) in the Thomas-Fermi-von-Weizs\"{a}cker
(TFvW) approximation \cite{tfvw_atom,tfvw_hydro}. We treat
image-charge interactions within the weighted density
approximation (WDA) \cite{ashcroft_wda,wda_im} for the exchange
and correlation energy of interacting electrons. In this way we
obtain improved description of the electronic potential energy,
which is regular close to the surface and exhibits correct
image-charge behavior inside vacuum. In Sec. III we apply our
method to field emission from jellium metallic surfaces and
discuss numerical results. Sec. IV contains our main conclusions.
Unless otherwise stated, we use atomic units ($e=\hbar=m=1$)
further throughout this work.

\section{Theoretical model}

In hydrodynamical approximation, the electron gas is treated as a
classical fluid which is characterized at each point in space by
number density $n(\rs t)$ and velocity $\vs(\rs t)$ distributions.
Neglecting dissipative effects, the electron dynamics is governed
by the Euler's equation
\begin{equation}
m \left(\frac{\partial \vs}{\partial t} + \vs \cdot \nabla \vs
\right) =- \nabla \frac{\delta {\cal E}[n]}{\delta n}  ,
\label{velocity}
\end{equation}
supplemented by the continuity equation for the density
\begin{equation}
\frac{\partial n}{\partial t} + \nabla \cdot ( n \vs) = 0
\label{density}.
\end{equation}
Here ${\cal E}$ is the internal energy of a static charged fluid,
which we approximate within DFT
\begin{eqnarray}
& & {\cal E}[n(\rs t)]= \int d^3 \rs ~ n(\rs t) V_{{\rm ext}}(\rs
t) + \nonumber
\\
& & + \int d^3 \rs \left[ \frac{3}{10} (3 \pi^2)^{2/3} n^{5/3}(\rs
t) +  \frac{\lambda}{2} \left(\frac{1}{2} \frac{\nabla_{\rs} n(\rs
t)}{n(\rs t)}\right)^2 n(\rs t)
\right] + \nonumber \\
& & + \frac{1}{2} \int d^3 \rs \int d^3 \rs' n(\rs t)
\frac{1}{|\rs-\rs'|} n(\rs' t) + {\cal E}_{{\rm xc}}[n(\rs t)],
\end{eqnarray}
The first term gives the potential energy of the valence electrons
in the electric field of the substrate ionic cores superimposed on
an external uniform electric field applied to the surface. The
second and third term represent the quantum-mechanical kinetic
energy of non-interacting electrons in the TFvW approximation with
phenomenological density gradient parameter $\lambda$. The value
$\lambda=1/4$ is used throughout this work as it gives good
quantitative results for the ground-state workfunction and surface
energy \cite{tfvw_lam}. The last two terms are the repulsive
Coulombic interaction energy and the exchange-correlation energy
of the electrons, respectively. The exchange-correlation energy
\begin{equation}
\label{exc} {\cal E}_{{\rm xc}}[n]=\frac{1}{2}\int d^3 \rs ~n(\rs
t) \int d^3 \rs' \frac{\bar{n}_{{\rm xc}}[n;\rs,\rs']}{|\rs-\rs'|}
\end{equation}
is the Coulomb interaction energy between an electron at the point
$\rs$ with its surrounding charge deficiency hole near $\rs'$
\begin{equation}
\label{nxc} {\bar n}_{{\rm xc}}[n;\rs,\rs']=n(\rs' t)
(\bar{g}_{{\rm xc}}[n;\rs,\rs']-1).
\end{equation}
The exchange-correlation hole $\bar{n}_{{\rm xc}}$ is given in
terms of coupling-constant averaged electron-pair distribution
function
\begin{equation}
\bar{g}_{{\rm xc}}[n;\rs,\rs']=\int_0^{1} d \gamma ~g_{{\rm
xc}}[n, \gamma;\rs,\rs'],
\end{equation}
here $g_{{\rm xc}} [n,\gamma]$ is the distribution function for
electrons interacting via modified pairwise Coulomb potential
$\gamma / r_{12}$ of strength $\gamma$, but with one-particle
density held fixed at the physical density ($\gamma=1$) due to
interaction of electrons with compensating $\gamma$-dependent
external one-body potential \cite{xc}. By definition ${\bar
n}_{{\rm xc}}$ satisfies the sum rule
\begin{equation}
\label{sumrule} \int d^3 \rs'  {\bar n}_{{\rm xc}}[n;\rs,\rs']=-1.
\end{equation}
In the weighted-density-approximation \cite{ashcroft_wda,GJL_xc2},
we keep the one-particle density pre-factor in Eq.(\ref{nxc}) and
replace the exact pair-correlation function with that of a
homogeneous electron gas of an auxiliary density $\tilde{n}$, i.e.
\begin{equation}
\label{gxcheg} \bar{g}_{{\rm xc}}[n;\rs,\rs'] \approx
\bar{g}^{{\rm hom}}_{{\rm xc}}(\tilde{n}(\rs t),u)
\end{equation}
where $u=|\rs-\rs'|$ is the relative distance between pair of
electrons. The effective density $\tilde{n}$ is a weighted average
of the true electron density $n(\rs)$ over the local Fermi
wavelength $\lambda_F=2 \pi / (3 \pi^2 n)^{1/3}$. For regions of
slowly varying density, the pre-factor in Eq.(\ref{nxc}) can be
treated as constant and brought in front of the integral in
Eq.(\ref{exc}), resulting in LDA.  The weighted-density parameter
is chosen such that to satisfy the sum rule in Eq.(\ref{sumrule}).
A model for the pair correlation function of the homogeneous
electron gas is given by the ansatz \cite{wda_im}
\begin{equation}
\bar{g}^{{\rm hom}}_{{\rm xc}}(\tilde{n},u) -1 \approx
C(\tilde{n}) \left[1-\exp\{-r^5_0(\tilde{n})/u^5 \} \right],
\end{equation}
in terms of two density parameters $C$ and $r_0$. The long-range
tail $u^{-5}$ of the correlation factor ensures image-like
behavior of the potential energy of an electron at large distances
from the surface. The density functional in Eq.(\ref{nxc}) with
the approximation of Eq.(\ref{gxcheg}) is also exact in several
limiting cases: i) for the homogeneous electron gas, ii) for
one-electron systems, such as hydrogen, where it gives exact
cancellation of the electrostatic Coulomb self-interaction energy
and iii) for an atom, where it reproduces the correct asymptotic
form of the exchange-correlation energy density $\eps_{{\rm
xc}}[n,\rs] = -1/2 |\rs|$ far from the nucleus.

\begin{figure}[h]
\begin{center}
\includegraphics
{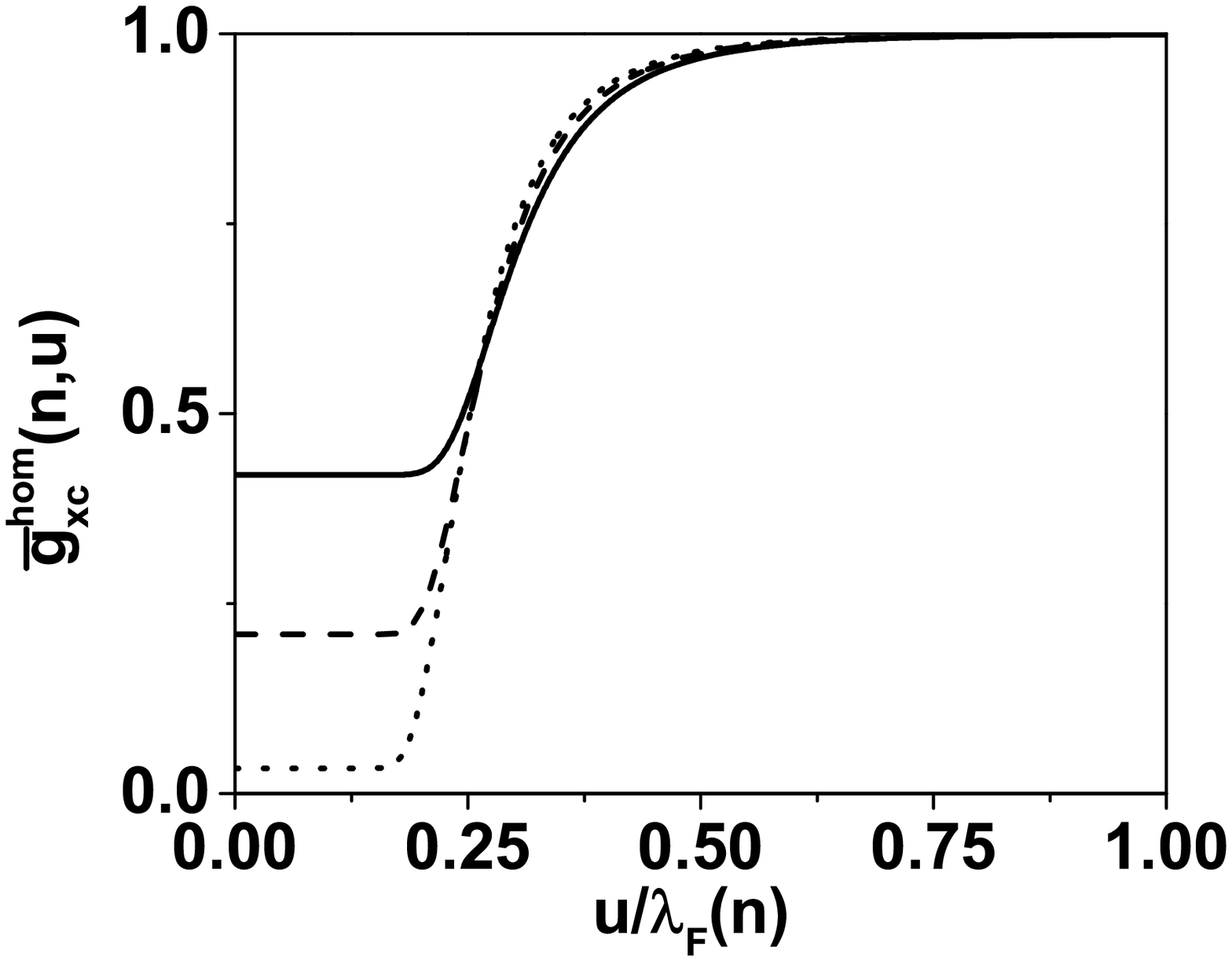}
\caption{Averaged electron-pair correlation function for the
homogeneous electron gas with density parameter $r_s=2$ (solid
line), $r_s=4$ (dashed line) and $r_s=6$ (dotted line). The
distance between the particles is given in Fermi wavelengths. }
\label{fig:gbar}
\end{center}
\end{figure}

At each point $\rs$, the parameters $C$ and $r_0$ are functions of
the weighted density $\tilde{n}$, their density dependence is
fixed by the satisfaction of the particle conservation sum rule
\begin{equation}
4 \pi \tilde{n} \int_0^{\infty} du u^2 [\bar{g}^{{\rm hom}}_{{\rm
xc}}(\tilde{n},u)-1]=-1
\end{equation}
and by the requirement that the exchange and correlation energy
per electron in an uniform system of  number density $\tilde{n}$
is
\begin{equation}
2 \pi \tilde{n} \int_0^{\infty} du u [\bar{g}^{{\rm hom}}_{{\rm
xc}}(\tilde{n},u)-1]=\eps_{{\rm xc}}(\tilde{n}).
\end{equation}
We approximate $\eps_{{\rm xc}}=[-0.458 /r_s-0.44 /(r_s+7.8)]$
with the sum of the Dirac exchange and the Wigner-corrected
correlation energy, where $r_s=(3/4 \pi \tilde{n})^{1/3}$ is the
Wigner-Seitz radius. The pair-correlation function is shown in
Fig. \ref{fig:gbar} for bulk Wigner-Seitz radii $r_s=2,4$ and $6$.
It represents the tendency of electrons to avoid one another at
small distances as a consequence of the short-range part of the
repulsive Coulomb interaction and due to the Pauli repulsion
between electrons in parallel spin component. Because
$\bar{g}^{{\rm hom}}_{{\rm xc}}$ tends to unity in a power-law
way, electrons become weakly correlated at distances $u
> \lambda_F/2$. A known drawback of this approximation is that the
correlation factor becomes negative for low density substrates
with $r_s > 6$, however it remains positive in the region of
metallic densities $1 \le r_s \le 6$.

\subsection{Field emission from jellium metallic
surfaces}

As an application, we investigate the steady-state solutions of
the hydrodynamical equations corresponding to field emission from
simple jellium metallic surfaces. Here we assume the velocity
distribution is described by an irrotational fluid flow ($ \nabla
\times \vs = {\bf 0}$). The  potential energy term $V_{{\rm ext}}$
represents the interaction of electrons with externally applied
electric field $\ec=-E \hat{z}$ superimposed on the electric field
of a neutralizing charge distribution of the substrate ionic cores
\begin{equation}
V_{{\rm ext}}(z)= -E z -\int d^3 \rs' \frac{
n_{+}(z')}{|\rs-\rs'|} + \langle \delta v \rangle_{{\rm ws}}
\theta(-z) ,
\end{equation}
where $n_+(z)=n_b \theta(-z)$ is the number density of the ionic
charge distribution, $\langle \delta v \rangle_{{\rm ws}}$ is a
pseudopotential correction to the ionic background potential,
which takes in an average way the difference between the total
pseudopotential of a discrete ion-lattice and the potential due to
the uniform positive background. The pseudopotential parameter
$\langle \delta v \rangle_{{\rm ws}}$ is choosen to make the
jellium mechanically stable \cite{jellium_perdew,jellium_diaz} and
does not cleave spontaneously. Due to the translation invariance
of the jellium electronic structure in the surface plane, the
number density and velocity distributions are functions of the
coordinate normal to the surface, $n=n(z)$ and $\vs(\rs)= v(z)
\hat{ {{\bf z}}}$. In this case, Eq.
(\ref{velocity})-(\ref{density}) expresses the conservation of the
current density $J= -e n v$  and the total energy per electron
\begin{equation}
\frac{J^2}{2 n^2} -\frac{\lambda}{4} \left[ \frac{n''}{n} -
\frac{1}{2} \left(\frac{n'}{n} \right)^2 \right] + \frac{1}{2} (3
\pi^2 n)^{2/3} +U_s  [n;z]= \mu, \label{Bernoulli}
\end{equation}
here $(n',n'')$ are the first and second derivatives of the
electron density, $\mu$ is the chemical potential and
\begin{equation}
U_s[n;z] = \phi[n;z]+ v_{{\rm xc}}[n;z]
\end{equation}
is a one-electron potential energy which is a superposition of the
electrostatic Hartree and effective exchange and correlation
energies, respectively. The Hartree potential $\phi$ satisfies the
Poisson's equation
\begin{equation}
\phi'' = -4 \pi [n(z)-n_{+}(z)]
\end{equation}
subject to the boundary condition $\phi'=-E$  at $z=z_{{\rm
max}}$. In WDA, the effective exchange-correlation potential is
given approximately by
\begin{eqnarray}
& & v_{{\rm xc}}[n;z] \approx \eps_{{\rm xc}}[n;z] +\int dz'
n(z') \omega_1(n_b;|z-z'|) + \nonumber \\
& & + \left( \eps_{{\rm xc}}(n_b) - n_b \frac{\eps_{{\rm
xc}}(n_b)}{d n_b} \right) \int dz' n(z') \omega_2(n_b;|z-z'|)
\end{eqnarray}
here
\begin{equation}
\eps_{{\rm xc}}[n;z]= \int dz' n(z') \omega_1(\tilde{n}(z);|z-z'|)
\end{equation}
is the exchange and correlation energy per electron, and the
functions $\omega_1(n,u)= \pi C(n)[r_0(n)
\gamma(4/5,r^5_0(n)/u^5)-u(1-\exp(-r^5_0(n)/u^5)]$ and
$\omega_2(n,u)= \pi C(n)[ r^2_0(n) \gamma(3/5,r^5_0(n)/u^5)-
u^2(1-\exp(-r^5_0(n)/u^5)]$ are given in terms of the incomplete
gamma function $\gamma(a,x)$. Inside vacuum, $v_{{\rm xc}}(z
\rightarrow \infty)=V_{{\rm im}}(z \rightarrow \infty)=-1/4z$
reproduces the correct classical image potential. Introducing the
density variable $M=\sqrt{n}$, the Bernoulli equation
Eq.(\ref{Bernoulli}) can be recast into a self-consistent Milne
equation \cite{pam}
\begin{equation}
\label{milne} M'' + \frac{2}{\lambda} \left[\mu-U_{{\rm
eff}}(z)\right] M= \frac{J^2}{\lambda M^3}
\end{equation}
in terms of an effective one-electron potential energy
\begin{equation}
U_{{\rm eff}}(z)=\frac{1}{2} [3 \pi^2 n(z)]^{2/3}+U_s(z) ,
\end{equation}
including the Thomas-Fermi kinetic pressure term and the
Coulomb-derived Hartree and exchange-correlation potential
energies.
For convenience we set the zero of energy to $\mu=0$. The boundary
condition for the charge density inside vacuum $z=z_{{\rm max}}$
can be written as $n \approx J/\sqrt{2[E(z-z_0)-W]}$ for some
constant $z_0$, which we further associate with the centroid of
the induced charge at surface in response to the external electric
field and $W$ is the workfunction. In the metal interior
$z=z_{{\rm min}}$, the electric field is fully screened by the
conduction electrons and the density approaches the bulk density
$n_b$, such that we impose two additional boundary conditions
$\phi'=0,M=\sqrt{n_b}$ for $z=z_{{\rm min}}$. The last two
conditions constitute a system of algebraic equations for the
integration constants $J$ and $z_0$, which were solved by the
Newton-Raphson method.  Eq.(\ref{milne}) was solved
self-consistently using the Bulirsch-Stoer method \cite{numrec}.
We use $z_{{\rm max}}= 5 \lambda_F$ and $z_{{\rm min}}=-2
\lambda_F$.

\section{Numerical result and discussion}

\subsection{Charge density distribution}

\begin{figure}[h]
\begin{center}
\includegraphics
{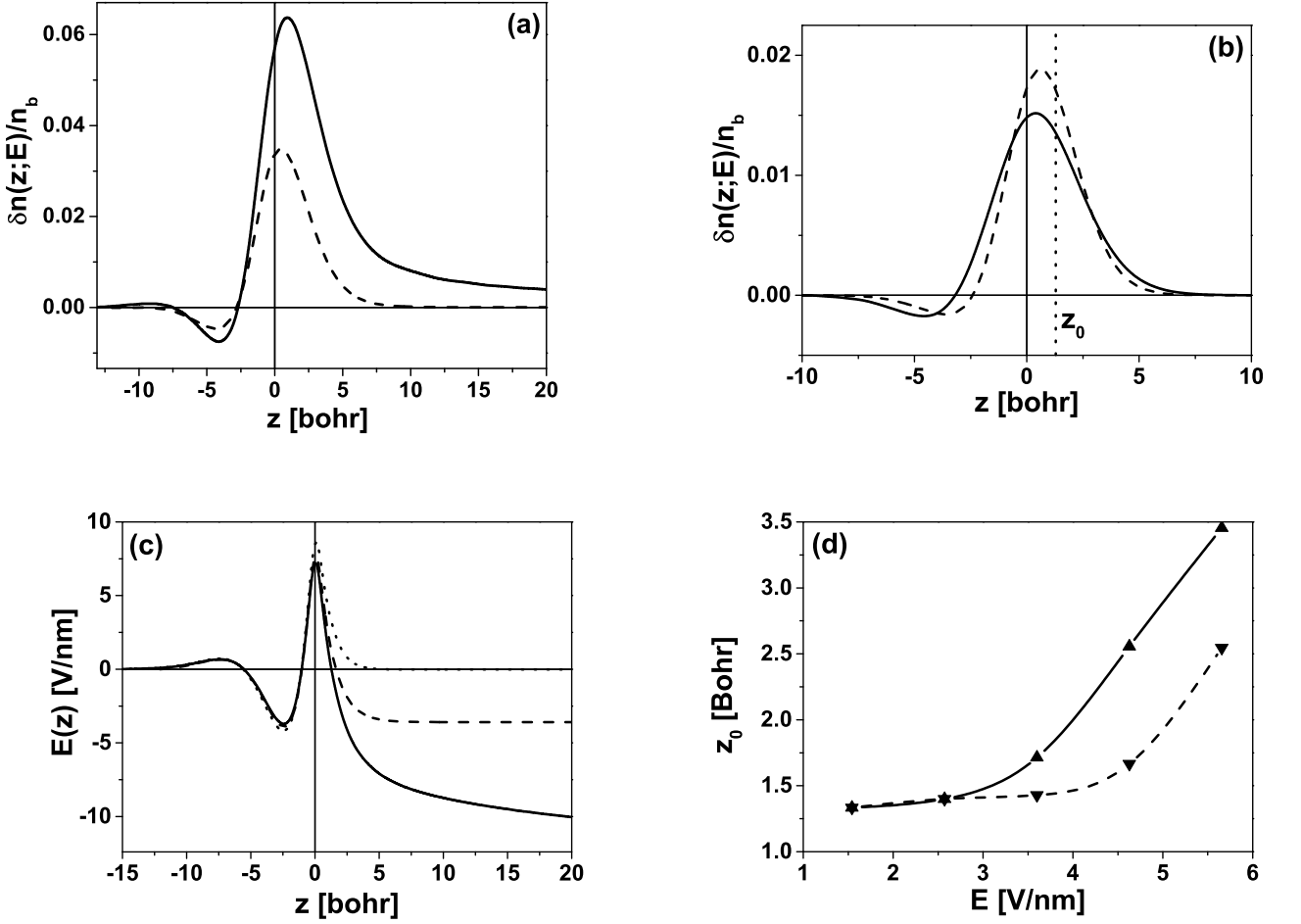}  \caption{
(a) Induced charge density $\delta n(z;E)/n_{{\rm
bulk}}=(n(z;E)-n_0(z))/n_{{\rm bulk}}$ normalized to the bulk
density as a function of the distance from the surface $z$. The
external electric field is $E=3.6$ V/nm (dashed line) and $E=12.3$
V/nm (solid line). (b) Induced screening charge density for
$E=1.5$ V /nm, dashed line (LDA) and solid line (WDA). The
centroid $z_0$ of the screening charge is indicated by the dashed
line. (c) Profile of the total electric field $E(z)$ for
externally applied electric field $E=3.6$ V/nm (dashed line),
$E=12.3$ V/nm (solid line) and the field-free result $E=0$ (dotted
line). (d) Field dependence of the screening parameter $z_0$ in
LDA (lower triangles) and  WDA (upper triangles). The distance $z$
in (a-c) is given in bohrs, and is measured relative to the
position of the jellium edge ($z=0$), indicated by the solid
vertical line. The Wigner-Seitz radius is $r_s=4$.
}\label{fig:ind}
\end{center}
\end{figure}

In Fig. \ref{fig:ind} (a), for $E=3.6$ V/nm and $E=12.3$ V/nm we
show the WDA induced charge density $\delta n(z;E) =
n(z;E)-n_0(z)$ relative to the equilibrium distribution for
$r_s=4$. For the lower electric field $E=3.6$ V/nm, the density
distribution is extended over the surface region and acts to
screen the applied electric field inside the metal, charge
transfer into vacuum is unlikely. In response to the increased
field strength $E=12.3$ V/nm, the screening charge at the surface
is enhanced, but the density distribution does not remain
localized and extends into vacuum due to tunneling transitions.
The resultant space-charge layer partially screens out the
external electric field inside vacuum and near the surface, as
Fig. \ref{fig:ind}(c) shows. Since the space-charge effect can be
neglected for weak electric fields $E \le 2$ V/nm, the induced
charge density profile is well characterized by its center-of-mass
\begin{equation}
z_0(E) = \frac{\int dz z \delta n(z;E)}{ \int dz \delta n(z;E) },
\end{equation}
which determines the effective location of the surface and also
serves as a reference plane for the classical image potential
\cite{LK_image}. The normalized LDA and WDA  screening charge
densities are shown in Fig.\ref{fig:ind}(b) for the lowest
electric field $E=1.5$ V/nm considered in this work. Also
indicated is the center of the induced charge, which is positioned
at a distance $z_0=1.27$ in front of the jellium edge for WDA and
at $z_0=1.32$ for LDA. However the WDA density is much more
diffuse, since the sharp density variations at the surface (which
are more pronounced in the LDA) are averaged out over the range
$r_0$ of the repulsive short-ranged electronic interactions. Since
the WDA density approaches much more slowly the uniform bulk
density in the metal interior, the external electric field is less
efficiently screened (cf. Fig.\ref{fig:ind}(c)).  In response to
increase of the field strength, the effective position of the
surface shifts toward vacuum as shown in Fig. \ref{fig:ind}(d).
That is because conduction electrons spill out at larger distances
inside vacuum due to enhanced penetration of the barrier, in good
qualitative agreement with the outward shift of $z_0$ for
negatively charged surfaces reported in Ref.\cite{chargedJe}. At
high $E\ge 10$ V/nm, the effective location of the surface is
difficult to determine due to the formation of negative space
charge layer inside vacuum (cf. Fig. \ref{fig:ind}(a)).

\begin{figure}[h]
\begin{center}
\includegraphics
{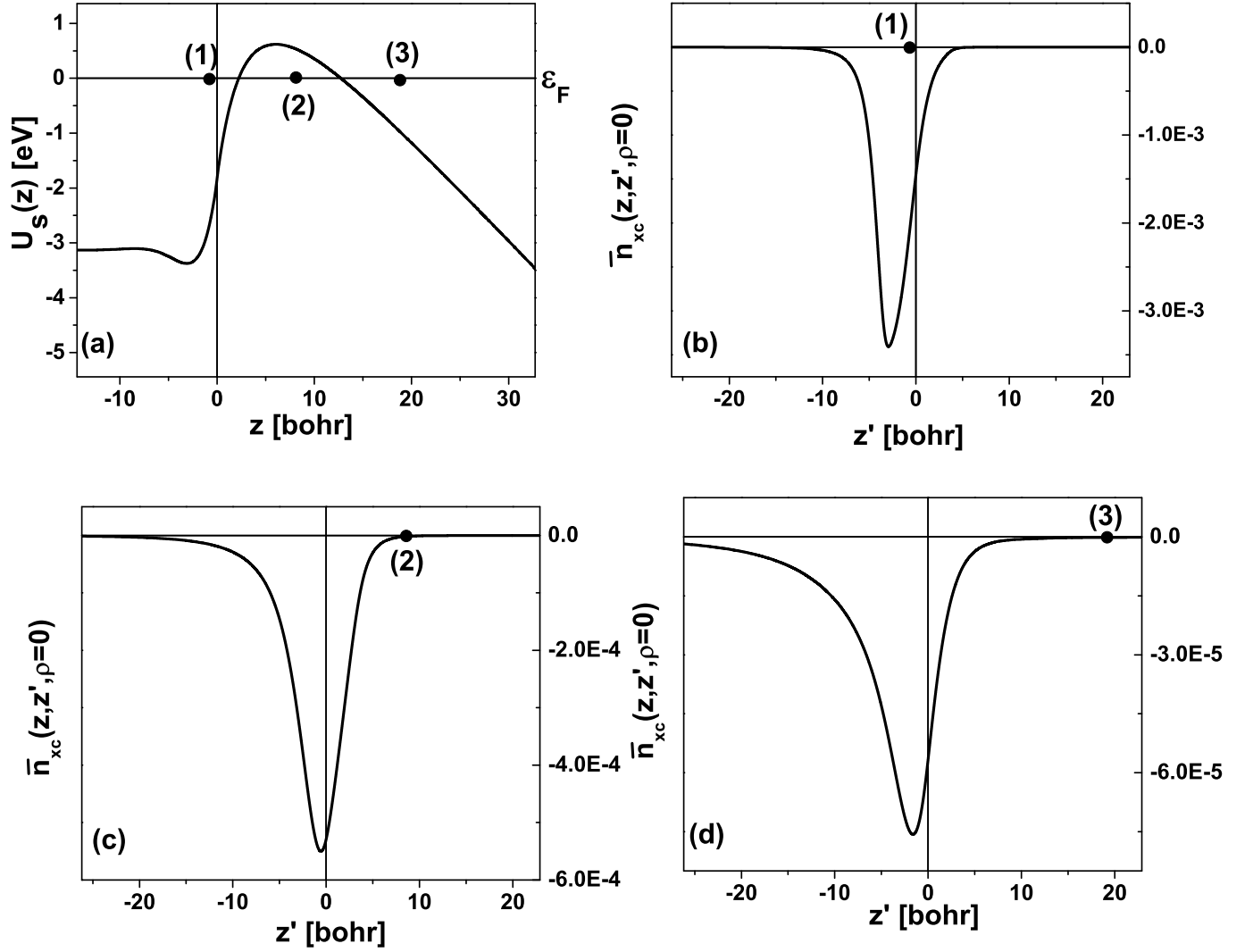}  \caption{
(a) Potential energy of an electron near jellium metallic surface
with $r_s=4$. The metal workfunction is $W_{{\rm WDA}} =2.65$ eV.
The position of the Fermi level is indicated by the solid vertical
line. The points (1-3) designate the location of an electron
inside the bulk, under the barrier and inside vacuum,
respectively. Fig. (b-d) give the exchange-correlation hole
$\bar{n}_{{\rm xc}}(z,z',\rho)$ about an electron at (1-3),
respectively. The lateral electron-hole separation is $\rho=0$.
The applied electric field is $E=3.6$ V/nm. The distances $z$ and
$z'$ in bohrs are measured relative to the  jellium edge,
indicated by the solid vertical line in Fig.(a-d).
}\label{fig:xch}
\end{center}
\end{figure}

Fig. \ref{fig:xch}(a) shows the effective one-electron potential
energy $U_s(z)$ for sodium metallic surface ($r_s=4$) subject to
an external electric field $E=3.6 $ V/nm. The various locations of
an electron inside the bulk (1), under the barrier (2) and inside
vacuum (3) are indicated. Fig. \ref{fig:xch}(b-d) give the
coupling-constant averaged exchange-correlation hole corresponding
to the different positions (1-3) in Fig. \ref{fig:xch}(a). In the
metal interior, cf. Fig.\ref{fig:xch}(b), the hole is centered on
the electron and the LDA approximation is adequate. When an
electron changes its position and crosses the surface by moving
outward under the barrier as shown in Fig.\ref{fig:xch}(c), the
hole does not follow and separates by spreading laterally along
the surface plane. The center of the hole does not remain
localized in the surface region and moves toward the bulk, as a
consequence of the weighted density approximation. The electron is
detached from the hole far inside vacuum as Fig.\ref{fig:xch}(d)
shows. The Coulomb attraction between the separated electron and
hole charges gives rise to the classical image force exerted on
the electron at large distances from the surface. Thus the
asymptotic potential energy becomes
\begin{equation}
U_{{\rm im}}(z \rightarrow \infty)=-\frac{1}{4 (z-z'_0)},
\end{equation}
where $z'_0$ specifies the position of the WDA image charge plane
\cite{CT_wda}
\begin{equation}
z'_0 =0.1143  \frac{ \eps_{{\rm xc}}(n_b)- n_b d \eps_{{\rm
xc}}(n_b)/ dn_b}{ \eps_{{\rm xc}}^2(n_b) }.
\end{equation}
This image plane does not depend on any detail of the surface
density distribution, thus $z'_0$ given by the WDA is artificially
fixed by the model for the pair correlation function of the
uniform electron gas. More importantly $z'_0$ is slightly shifted
inside the jellium and does not coincide with the centroid $z_0$
of the induced (negative) screening charge at the surface. Despite
this limitation of WDA, it is known that the screening response of
a jellium substrate subjected to external electric fields of
distinguishable test charges is characterized by $z_0$ which is
different from $z'_0$ experienced by electrons. For Al(111)
surfaces, the position of the image charge plane is closer to the
jellium edge than that predicted from classical response to
external electric fields \cite{GW}. The importance of field
emission in testing the equality between the parameters $z_0$ and
$z'_0$ has also been pointed in \cite{LK_image}. In Tab. I, we
give values for the charge centroid $z_0$ and the WDA image-plane
$z'_0$ compared to other results. Our $z'_0$ slightly
overestimates the analytic result of Ref.\cite{CT_wda}, but
otherwise gives negative $z'_0$. In contrast, the centroid of the
induced screening charge $z_0$ lies in front of the surface and
shifts toward the jellium edge with increase of the bulk density.
Such variation of $z_0$ exhibits an opposite trend to LDA result
for ordinary jellium \cite{SHG}. This difference stems from the
inclusion of the discrete ion-lattice effects in the stabilized
jellium model, which in turn modify the screening properties of
the metal (cf. Ref. \cite{Kiejna}).

\begin{table}[h]
\caption{Centroid of the induced screening charge $z_0$ in bohrs
for simple jellium surfaces subject to weak electric field $E=1.5$
V/nm, and position of the WDA image plane $z'_0$. The centroid
$z_0$ from Ref.\cite{Kiejna} and Ref.\cite{SHG} are results of a
LDA calculation in the stabilized and ordinary jellium models,
respectively. Ref.\cite{CT_wda} gives analytic result for the
location of the WDA image charge plane $z'_0$. }
\centering 
\begin{tabular}{l c c c c c}
\hline\hline
 $r_s$ &  $z'_0$  & $z'_0$ Ref.\cite{CT_wda} & $z_0$ & $z_0$ Ref.\cite{Kiejna} & $z_0$ Ref.\cite{SHG}   \\ [0.5ex]
\hline %
2  & -0.08 &   -0.29  &  0.7      & 0.97  & 1.57 \\
3  & -0.22 &   -0.42  &  1.2     & 1.05  & 1.35 \\
4  & -0.34  &  -0.54  & 1.3  & 1.19  & 1.25 \\
 \hline
  \end{tabular}
\label{tab:field}
\end{table}

\subsection{One-electron potentials}

\begin{figure}[h]
\begin{center}
\includegraphics
{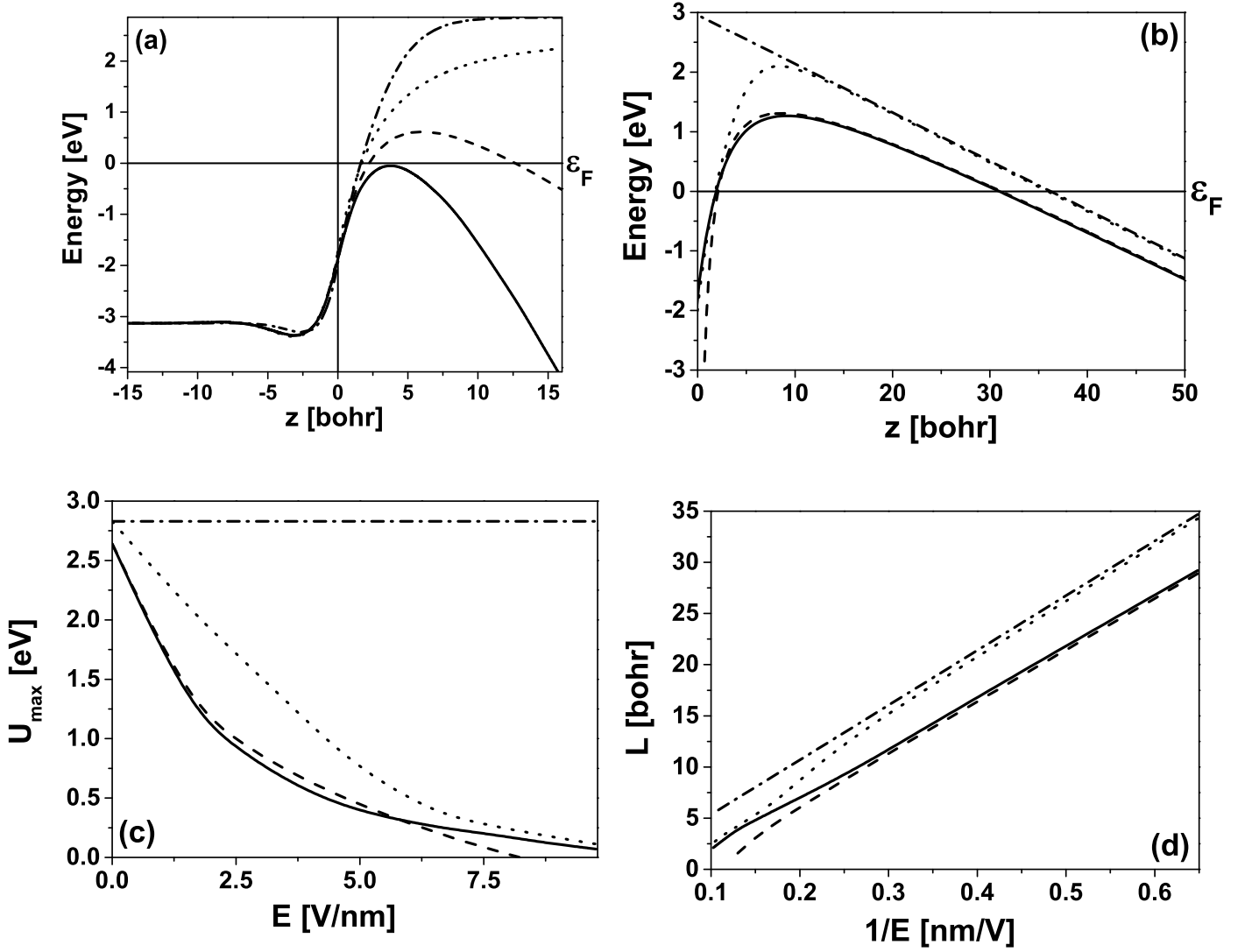}  \caption{
(a) Potential energy of an electron near jellium metallic surface
with $r_s=4$. The field-free potential energy is given by the
dashed-dotted (LDA) and the dotted line (WDA). The horizontal
solid line gives the position of the Fermi energy level. The
dashed line gives $U_s(z)$ in  WDA for external electric field
$E=3.6$ V/nm and the solid line for $E=12.3$ V/nm. (b) Surface
barrier experienced by electrons for $E=1.5$ V/nm and $r_s=4$. The
distance $z$ is measured relative to the jellium edge $z=0$.
Fig.(c) gives the surface barrier height (in eV) for an electron
near Fermi level as a function of the electric field. Fig. (d)
gives the full width of the surface barrier (in bohrs) for an
electron at the Fermi level as a function of the inverse field
strength. In. Fig.(b-d), the Fowler-Nordheim results are given by
the dashed-dotted lines, the modified Murphy-Good ones by the
dashed lines, the LDA  and WDA ones by the dotted and solid lines,
respectively.
}\label{fig:veff4}
\end{center}
\end{figure}

Fig.\ref{fig:veff4}(a) shows the field-dependence of the
one-electron potential energy $U_s(z;E)$ derived from the density
distributions in Fig. \ref{fig:ind}(a). For comparison, the
field-free potentials ($E=0$) are also shown. The zero-field WDA
potential energy interpolates smoothly between the bulk LDA energy
for large negative $z$ and the static image potential energy $-1/4
z$ for large distances to the surface, as shown by the dotted line
in Fig.\ref{fig:veff4}(a). In contrast the LDA potential energy
falls exponentially with the distance, since it does not represent
the non-local effect of separation of an electron from its
exchange-correlation hole after the electron has moved inside
vacuum. In response to moderate external electric field $E=3.6$
V/nm, the induced negative charge density at the surface lowers
the height of the surface barrier and the width of the
under-barrier region at the Fermi level becomes comparable to
$\lambda_F$, which increases the probability for vacuum tunneling.
The increased field strength continues to lower the barrier height
and reduce the width of the surface barrier. For $E=12.3$ V/nm the
top of the barrier (the Schottky saddle) moves below the position
of the Fermi level to enable classically-allowed over-barrier
transition.

\begin{figure}[h]
\begin{center}
\includegraphics
{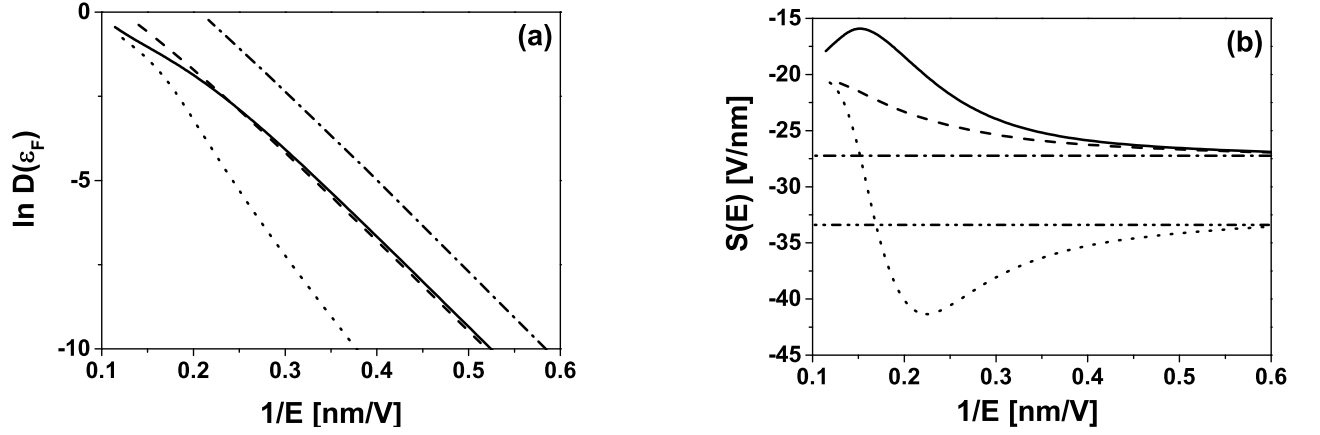}  \caption{
(a) Logarithm of the WKB barrier transmission coefficient for an
electron at the Fermi level as a function of the inverse field
strength (in nm/V) for $r_s=4$. Original Murphy-Good model
(dashed-dotted line), modified Murphy-Good model (dashed line),
WDA (solid line) and LDA (dotted line). (b) Slope of the
transmission coefficient (in V/nm) as a function of the inverse
field strength in (nm/V). The dashed-dotted line gives the
field-independent Fowler-Nordheim slope corresponding to the
workfunction $W_{{\rm WDA}}=2.65$ eV, the double dotted dashed
line corresponds to the Fowler-Nordheim slope with LDA
workfunction $W_{{\rm LDA}}=2.85$ eV. Solid-line: WDA,
dotted-line: LDA, dashed-line: modified MG. }\label{fig:wkb}
\end{center}
\end{figure}

In Fig.\ref{fig:veff4}(b), we plot the LDA and WDA surface
barriers for $E=1.5$ V/nm and $r_s=4$. For comparison the FN
barrier $U_{{\rm FN}}(z;E)=W-E(z-z_0)$ and the modified
Murphy-Good barrier are also shown. Here a modified MG barrier is
introduced to incorporate the difference between the first moments
of the induced screening-charge and image-charge distributions
\begin{equation}
U_{{\rm MG}}(z;E)=W- E (z-z_0)-\frac{1}{4 (z-z'_0)} ,
\label{MGpot}
\end{equation}
If we choose $z'_0=0$ as an effective position of the surface, the
potential energy in Eq.(\ref{MGpot}) becomes identical with the
classical MG result $U=\tilde{W}-Ez-1/4 z$ but with upshifted
workfunction $\tilde{W}=W+Es$, where $s=z_0-z'_0
>0$ is the separation between the centroid of the induced
negative charge density and the image charge plane experienced by
electrons leaving the surface. It is  worth to note that the
effect of the workfunction increase is the same as if the surface
is covered with electronegative adsorbates, such that the
contribution of the induced dipole layer increases the substrate
workfunction. The effective potential energy of Eq.(\ref{MGpot})
reproduces well the WDA result up to very small distances as
Fig.\ref{fig:veff4}(b) makes evident. In contrast, the LDA barrier
exhibits only weakly reduced height and width due to neglect of
image charge interaction, and thus follows more closely the
triangular Fowler-Nordheim barrier. The change of the surface
barrier height and width due to change of the external electric
field is shown in Fig.\ref{fig:veff4}(c-d). In the modified MG
model, the reduced barrier height for electrons near Fermi level
is
\begin{equation}
U_{{\rm max}}=W -\sqrt{E} + Es.
\end{equation}
As Fig.\ref{fig:veff4}(c) demonstrates, the modified MG barrier
height is nearly indistinguishable from the WDA result up to $7$
V/nm. Above $E > 7$ V/nm, the WDA barrier height reduces much more
slowly due to build up of the vacuum space charge layer, where
self-electric fields of mobile electrons oppose the external
field, such that the total electric field is weakened at the
surface. In contrast, the LDA barrier felt by electrons at Fermi
level is much higher at low electric fields and decreases linearly
with the increased field strength $E$, i.e.
\begin{equation}
U^{({{\rm LDA}})}_{{\rm max}} \approx W(1-E/E_d) , \quad E < E_d
\end{equation}
where fit to the LDA barrier height gives $E_d \approx 6$ V/nm.
The differences between the LDA and WDA barrier heights diminishes
at high $E \approx 10$ V/nm, when the Schottky saddle has moved
below the position of the Fermi level and electron transfer
becomes classically allowed. Below threshold $E < E_d$, the LDA
barrier breadth follows quantitatively the prediction of the
Fowler-Nordheim model $L_{{\rm LDA}}=W/E$ as shown in
Fig.\ref{fig:veff4}(d). Similarly the WDA barrier breadth follows
closely the modified MG result $L_{{\rm WDA}}=(\tilde{W}/E)
\sqrt{1-\tilde{y}^2}$, where $\tilde{y}=\sqrt{E}/ \tilde{W}$ is
the modified Nordheim parameter given  in terms of the square-root
field strength and the effective workfunction $\tilde{W}$. At
these relatively weak electric electric fields $E < 7$ V/nm, the
LDA barrier breadth is thicker than the WDA breadth by nearly $5$
bohr radii, such that electron tunneling is much more likely in
WDA. Using these parameters, the barrier transmission coefficient
can be approximated by $D \approx \exp(-\sqrt{2 U_{{\rm max}}}
L)$, after assuming that the multiplicative pre-exponential factor
of the tunneling rate changes negligibly with $E$. For tunneling
out of the LDA barrier we get
\begin{equation}
D_{{\rm LDA}} \approx \exp \left(-\sqrt{2} \frac{W^{3/2}}{E}
\sqrt{1-E/E_d} \right),
\end{equation}
and hence the screening response of conduction electrons is
expressed by a correction factor $\sqrt{1-E/E_d}$ to the
Fowler-Nordheim exponent. The slope of the LDA tunneling rate
(below threshold) is given by
\begin{equation}
S_{{\rm LDA}} = -E^2 \frac{\partial}{\partial E} \ln D_{{\rm
LDA}}(E) = - \sqrt{2} W^{3/2} \frac{1 - E/2 E_d}{\sqrt{1-E/E_d}} ,
\quad E < E_d
\end{equation}
so that the LDA slope is steeper relative to the field-independent
Fowler-Nordheim slope $S_{{\rm FN}}= -\sqrt{2} W^{3/2}$, when $E
\ll E_d$ the LDA slope converges $S_{{\rm LDA}} \rightarrow
S_{{\rm FN}}$. Quite similarly, the WDA result
\begin{equation}
D_{{\rm WDA}} \approx \exp \left(-\sqrt{2}
\frac{\tilde{W}^{3/2}}{E} f(\tilde{y}) \right) \label{MGmod}
\end{equation}
is given in in terms of slowly-varying barrier reduction factor
$f(\tilde{y})=(1-\tilde{y})\sqrt{1+\tilde{y}}$, which in contrast
to LDA is primary due to self-image charge screening. The
corresponding slope $ S_{{\rm WDA}} = - \sqrt{2} W^{3/2}
g(\tilde{y}), \quad \tilde{y} < 1$ with
$g(\tilde{y})=f(\tilde{y})(1-3eEs/2 \tilde{W})- (1-2
eEs/\tilde{W}) \tilde{y}/2 df / d\tilde{y}$, since $g < 1$ for
$\tilde{y} <1$, the WDA slope is flatter relative to $S_{{\rm
FN}}$. Thus for relatively weak fields, the slope of the WDA
emission rate is also flatter than the LDA one. However for
sufficiently weak fields $E \le 1$ V/nm, the effects of the
barrier reduction factors are weakened, hence the slopes $S_{{\rm
LDA}}$ and $S_{{\rm WDA}}$ should merge into the FN slope. But
since the LDA workfunction is increased by $0.2$ eV relative to
the WDA result (cf. Tab. II for $r_s=4$), the difference of the
slopes $S_{{\rm WDA}}-S_{{\rm LDA}} \approx \sqrt{2}
(W^{3/2}_{{\rm WDA}}-W^{3/2}_{{\rm LDA}})$ depends only on the
difference of the model workfunctions. In order to verify such a
qualitative estimate, in Fig.\ref{fig:wkb}(a-b) we plot the WKB
emission rate and its slope for an electron at the Fermi level,
\begin{equation}
D_{{\rm WKB}}(\epsilon_F) = \exp\left(-2 \int_{z_1}^{z_2} dz
\sqrt{2 U_s(z;E)} \right),
\end{equation}
where $z_1$ and $z_2$ are the inner and outer turning points of
the potential, for comparison the ordinary and modified MG results
are also shown. The ordinary MG result is obtained from
Eq.(\ref{MGpot}) by setting $s=0$, i.e. neglecting finite
differences between classical and the WDA image-charge planes. As
\ref{fig:wkb}(a) demonstrates, the modified MG and WDA emission
rates are indistinguishable up to $E=5$ V/nm, when the WDA
emission becomes slower relative to the modified MG result due to
the space charge effect. In contrast field emission in LDA is
highly unlikely, as it suppressed by more than one order of
magnitude relative to the WDA result. The ordinary MG emission
rate is the highest and overestimates the WDA result by a factor
of 5. To understand this enhancement, Taylor expansion of the
exponent in Eq.(\ref{MGmod}) for sufficiently weak fields $Es \ll
W$ results in $D_{{\rm 0}}=e^{2 \kappa s} D_{{\rm MG}}$, where
$\kappa=\sqrt{2 W}$ and $D_0$ is the ordinary MG emission rate.
For $r_s=4$, we have $2 \kappa=0.88$ and $s=1.8$, thus $\exp(2
\kappa s) \approx 5$, thereby reducing the modified MG emission
rate relative to the ordinary MG result by such a factor due to
weak field-induced increment of the substrate work-function by
$\Delta W = Es \approx 0.01$ eV.

\subsection{Current densities}

\begin{figure}[h]
\begin{center}
\includegraphics
{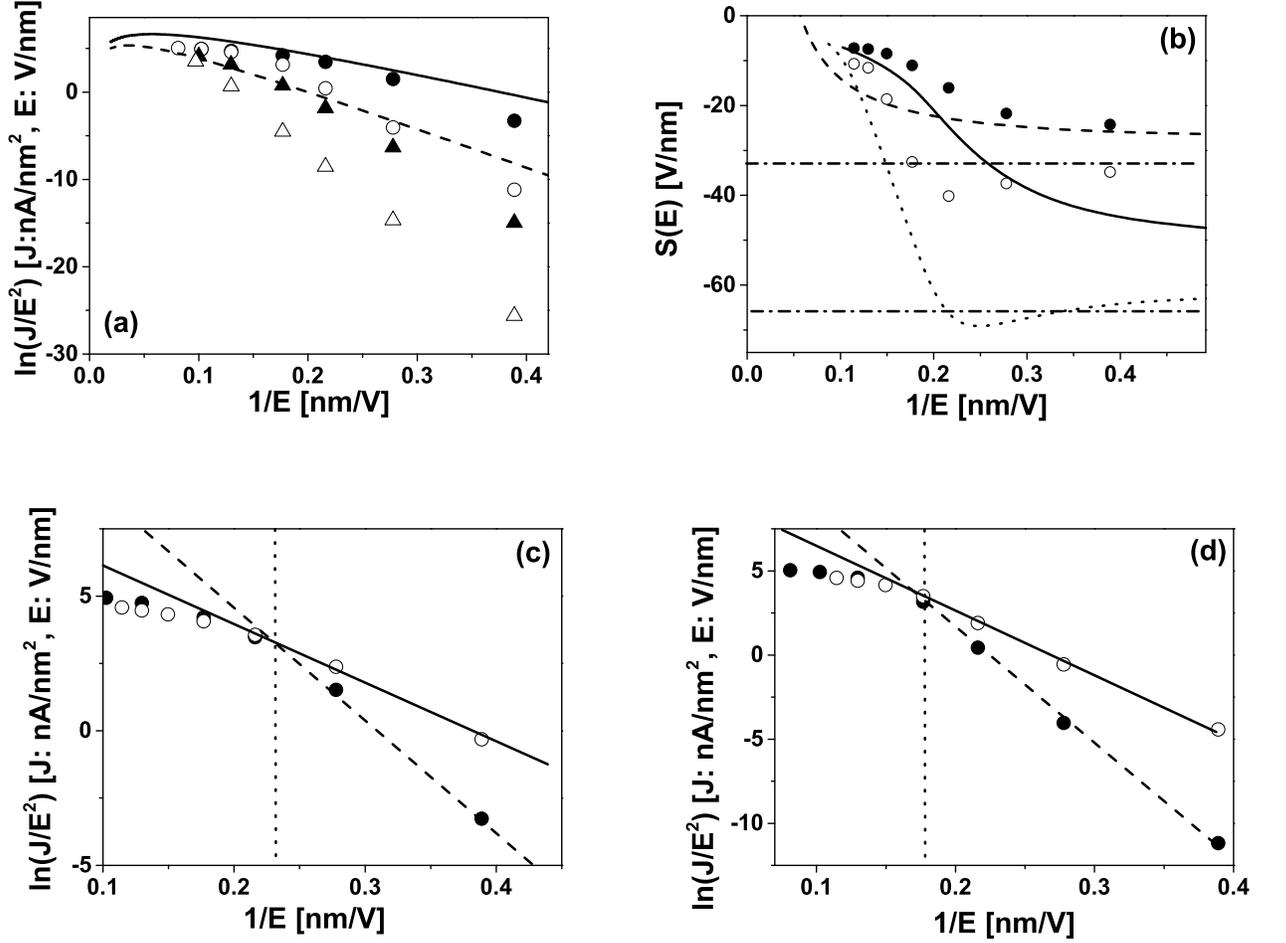}  \caption{
(a) Fowler-Nordheim plot calculated in the TFvW and modified MG
approximations. Open and filled circles give results in LDA and
WDA for $r_s=4$, respectively. Filled and open upward pointing
triangles give WDA and LDA results for $r_s=2$, respectively. For
$r_s=4$, the dashed and the solid lines give result of the
modified MG model with $r_s=4$ and $r_s=2$, respectively. The
current density $J$ is measured in [nA/nm$^2$] and the electric
field $E$ in [V/nm]. Fig. (b) Field-dependence of the slope of the
FN plot. Solid line gives the TFvW-WDA slope, dotted line -
TFvW-LDA slope, dashed line - modified MG slope, dashed-dotted
line gives the field-independent Fowler-Nordheim slope $S_{{\rm
FN}}$ with LDA workfunction. The double-dotted dashed line gives
the shifted Fowler-Nordheim slope $S(\lambda=1/4)=2 S_{{\rm FN}}$.
The empty and filled circles give WDA and LDA FN slopes in the WKB
approximation. Fig.(c) compares the FN plots in the TFvW and WKB
approximations for emission from the WDA surface barrier with
$r_s=4$. Fig.(d) the same as in Fig.(c), but for the emission from
the LDA barrier. The dashed and solid straight lines are fits of
the numerical result to an exponential formula Eq.(\ref{fit}). The
dotted vertical line designates the position of the threshold
field strength $E_d$, above which the emission mechanism changes
to space-charge limited field emission.} \label{fig:FNplot}
\end{center}
\end{figure}

In Fig.\ref{fig:FNplot}(a) we plot $\ln(J/E^2)$ as a function of
$1/E$ for aluminum ($r_s=2$) and sodium ($r_s=4$) surfaces. For
weak fields, the overall trend of decrease of $J$ with the
increase of the substrate work-function $W$ is clearly exhibited
in the TFvW model. For the lower screening radius $r_s=2$, the LDA
and WDA-FN plots are approximately linear functions of $1/E$, in
good qualitative agreement with the modified MG result. A
quantitative difference stems from the steeper slope of the WDA
tunneling rate relative to the MG result. This difference is a
direct consequence of the TFvW approximation for the kinetic
energy, which gives density distributions that are less extended
into vacuum, because the inverse decay length of the TFvW charge
density $\kappa(\lambda=1/4)=\sqrt{2 W / \lambda}$ is two times
larger than the equilibrium inverse decay length $\sqrt{2 W}$.
Similar conclusion can be drawn for the lower density jellium with
$r_s=4$. Noticeably, however the LDA rates exhibit even steeper
slope relative to the WDA result: for $E=3.6$ V/nm and $r_s=4$,
the LDA current density $J=2 \times 10^4 {{\rm A/cm^2}}$, while
the WDA result $J=6 \times 10^6 {{\rm A/cm^2}}$ is larger by two
orders of magnitude, as a consequence of the decreased WDA
workfunction.

For the lower density jellium $r_s=4$ and electric field $E
> 5$ V/nm, the slope of our FN plot is flatter than that at lower
$E$. The non-linearity of the FN plot agrees with the result
reported in Ref.\cite{Gohda}, which was attributed to combination
of several factors, such as finite energy width of the
distribution of emitted electrons, finite probability for
overbarrier reflection at the surface and the space charge effect.
As our LDA and WDA results become indistinguishable at high $E$,
this effect is insensitive to the description of electron
correlations. This non-linearity in the FN plot is due to build up
of the space charge layer inside vacuum (cf.
Fig.\ref{fig:ind}(a,c)) that partially screens out the applied
field inside vacuum, which tends to drive back electrons into the
metal. This effect flattens out the slope of the FN plot and
limits the increase of the current with the increased field
strength, consistent with the classical Child-Langmuir law
\cite{Child,Langmuir}.

To analyze this further, in Fig.\ref{fig:FNplot}(b), we plot the
slope $S(E)=-E^2 \partial_E \ln[J(E)/E^2]$ of the FN plot as a
function of the inverse field strength $1/E$ for $r_s=4$. For
comparison, the modified MG and FN slopes are also shown. At low
$E \sim 2 $ V/nm, $S_{{\rm LDA}}$ and $S_{{\rm WDA}}$ are slowly
varying functions of $E$, but are steeper as compared to the MG
and FN curves, because of the reduced density gradient
$\lambda=1/4$ in the TFvW approximation (as noted above). The LDA
and WDA slopes converge to the shifted Fowler-Nordheim slope
$S_{{\rm FN}}(\lambda=1/4)=-8/3 \sqrt{2} W^{3/2}$ for low $E$. The
WDA slope flattens with the increase of $E$ and continuously
crosses over to the Child-Langmuir behavior for $E > 7$ V/nm. In
contrast, the LDA slope steepens  with the increase of $E$,
attains a minimum near $E=4$ V/nm and flattens out to merge with
the WDA rate near $E=7$ V/nm.  The WKB result for the
electron-emission from the LDA and WDA surface barriers exhibits
qualitatively similar trend. But due to the increased decay length
of the electron density inside vacuum, the WKB slopes are shifted
upward relative to the TFvW slopes at low $E$, and converge to the
Fowler-Nordheim slope $S_{{\rm FN}}(\lambda=1)=-4/3 \sqrt{2}
W^{3/2}$. At higher $E >7$ V/nm, the WKB slopes are nearly
indistinguishable from the TFvW ones. Thus the quantitative
difference between TFvW and WKB rates is solely due to the reduced
density gradient $\lambda=1/4$, which affects electron spill out
inside vacuum.

Thus for relatively weak electric fields below some
surface-specific threshold field strength $E_d(r_s)$, the TFvW
result follows qualitatively the Fowler-Nordheim model in terms of
the exponential formula
\begin{equation}
\label{fit} J / E^2 = P \exp(S/E).
\end{equation}
The density dependent parameters $P,S$ and $E_d$ were extracted
from fit of the numerical result to Eq.(\ref{fit}) and given in
Tab. II. The TFvW slope parameter $S\approx -8/3 \sqrt{2} W^{3/2}$
coincides with the shifted Fowler-Nordheim slope $2 S_{{\rm
FN}}(\lambda=1)$, but the TFvW pre-factors exceed the MG and FN
results by few orders of magnitude.  For comparison, the
corresponding WKB slope and pre-exponential factors are also given
in Tab.II, the WKB slopes are flatter and tend to agree
qualitatively with the ordinary Fowler-Nordheim result.
Fig.\ref{fig:FNplot}(c-d) shows comparison of the WKB and TFvW
current densities for the LDA and WDA barriers. Above the
threshold $E_d$ these current densities are almost
indistinguishable, so the high-field part of the FN plot is
unaffected by the approximation for noninteracting kinetic energy.
The effect of the TFvW kinetic energy is to steepen the LDA and
WDA slopes relative to the WKB result just below the threshold
$E_d$. As a consequence the WKB pre-exponential factors are
reduced and move closer to the values predicted by the classical
FN and modified MG results. In both cases, the lowered WDA
workfunction relative to LDA is an essential factor in increasing
the emission current.

\begin{table}[h]
\caption{Work function $W$ in eV and slope $S$ of the FN plot in
V/nm in the TFvW ($\lambda=1/4$), FN, modified MG models for
different bulk Wigner-Seitz radii $r_s$ in bohrs, the last column
gives a pre-exponential factor $P$ (in A/V$^2$) in Eq.
(\ref{fit}). For comparison with TFvW, the WKB emission rates from
the LDA and WDA surface barriers are also shown. LDA and WDA
designate different models for the exchange and correlation energy
used throughout this work. The parameters $P$ and $S$ derived from
fit of our numerical data in Fig.(\ref{fig:FNplot}a) to Eq.
(\ref{fit}) for applied electric fields $E$ below a threshold
field strength $E_d$, above which the emission mechanism switches
to space-charge limited field emission.  The notation
$a[b]=a\times 10^b$ is used.}
\centering 
\begin{tabular}{l c c c c c c }
\hline\hline
 $r_s$ &   & $W$ & $E_d$ & $S$ & $P$  \\ [0.5ex]
\hline %
2  & &  & & & \\
  & TFvW-LDA  & 4.01 & 19 & -100 &  5.3[-4]  \\
  & TFvW-WDA  & 3.62 & 17 & -81 & 1.1[-2]   \\
  & FN   & 4.01 & & -55  & 6.3[-7]  \\
  & MG   & 3.62 & & -44  & 6.5[-6]   \\
  & WKB-LDA & 4.01 & 19 & -55  & 7.8[-4]  \\
  & WKB-WDA & 3.62 & 17 & -37  & 3.1[-4]  \\
4  & &  & & & \\
 &  TFvW-LDA  & 2.85 & 5.0 & -64 & 1.1[-3] \\
  & TFvW-WDA  & 2.65 & 4.5 & -48 & 3.4[-3] \\
  & FN   & 2.85 &  & -33  & 1.0[-6] \\
  & MG   & 2.65 &  & -26  & 1.7[-4]  \\
  & WKB-LDA  & 2.85 & 5.0 &  -37 & 2.1[-4]  \\
  & WKB-WDA  & 2.65 & 4.5 &  -23 & 5.2[-5] \\
 \hline
  \end{tabular}
\label{tab:field}
\end{table}

\section{Conclusion}

We have evaluated the field emission current from jellium metallic
surfaces in the framework of quasi-static
Thomas-Fermi-von-Weizs\"{a}cker model. Our result deviates from
the Fowler-Nordheim model for strong electric fields due to space
charge effects. In the tunneling regime we find the emission
mechanism depends sensitively on the properties of the surface
electronic structure and the approximations for kinetic and
exchange-correlation energies, which in turn affect the substrate
workfunction. As a consequence, we find that the conventional
local-density-approximation underestimates the emission current by
few orders of magnitude relative to the result of the
weighted-density-approximation. We also find that the WDA emission
current is decreased relative to the classical Murphy-Good result
due to effective increase of the substrate workfunction by
$0.01-0.02$ eV, which is due to the relative displacement between
the WDA image charge plane experienced by electrons leaving the
surface and the centroid of the induced screening charge density
at the surface.

\section*{Acknowledgments}

The authors acknowledge financial support by the
ANR-12-BS09-0013-02 grant.

\newpage

\end{document}